

\magnification=\magstephalf
\hsize=13.0 true cm
\vsize=19 true cm
\hoffset=1.50 true cm
\voffset=2.0 true cm

\abovedisplayskip=12pt plus 3pt minus 3pt
\belowdisplayskip=12pt plus 3pt minus 3pt
\parindent=2em


\font\sixrm=cmr6
\font\eightrm=cmr8
\font\ninerm=cmr9

\font\sixi=cmmi6
\font\eighti=cmmi8
\font\ninei=cmmi9

\font\sixsy=cmsy6
\font\eightsy=cmsy8
\font\ninesy=cmsy9

\font\sixbf=cmbx6
\font\eightbf=cmbx8
\font\ninebf=cmbx9

\font\eightit=cmti8
\font\nineit=cmti9

\font\eightsl=cmsl8
\font\ninesl=cmsl9

\font\sixss=cmss8 at 8 true pt
\font\sevenss=cmss9 at 9 true pt
\font\eightss=cmss8
\font\niness=cmss9
\font\tenss=cmss10

 at 12 true pt
\font\bigbf=cmbx10 at 12 true pt

\catcode`@=11
\newfam\ssfam

\def\tenpoint{\def\rm{\fam0\tenrm}%
    \textfont0=\tenrm \scriptfont0=\sevenrm \scriptscriptfont0=\fiverm
    \textfont1=\teni  \scriptfont1=\seveni  \scriptscriptfont1=\fivei
    \textfont2=\tensy \scriptfont2=\sevensy \scriptscriptfont2=\fivesy
    \textfont3=\tenex \scriptfont3=\tenex   \scriptscriptfont3=\tenex
    \textfont\itfam=\tenit                  \def\it{\fam\itfam\tenit}%
    \textfont\slfam=\tensl                  \def\sl{\fam\slfam\tensl}%
    \textfont\bffam=\tenbf \scriptfont\bffam=\sevenbf
    \scriptscriptfont\bffam=\fivebf
                                            \def\bf{\fam\bffam\tenbf}%
    \textfont\ssfam=\tenss \scriptfont\ssfam=\sevenss
    \scriptscriptfont\ssfam=\sevenss
                                            \def\ss{\fam\ssfam\tenss}%
    \normalbaselineskip=13pt
    \setbox\strutbox=\hbox{\vrule height8.5pt depth3.5pt width0pt}%
    \let\big=\tenbig
    \normalbaselines\rm}

\def\ninepoint{\def\rm{\fam0\ninerm}%
    \textfont0=\ninerm      \scriptfont0=\sixrm
                            \scriptscriptfont0=\fiverm
    \textfont1=\ninei       \scriptfont1=\sixi
                            \scriptscriptfont1=\fivei
    \textfont2=\ninesy      \scriptfont2=\sixsy
                            \scriptscriptfont2=\fivesy
    \textfont3=\tenex       \scriptfont3=\tenex
                            \scriptscriptfont3=\tenex
    \textfont\itfam=\nineit \def\it{\fam\itfam\nineit}%
    \textfont\slfam=\ninesl \def\sl{\fam\slfam\ninesl}%
    \textfont\bffam=\ninebf \scriptfont\bffam=\sixbf
                            \scriptscriptfont\bffam=\fivebf
                            \def\bf{\fam\bffam\ninebf}%
    \textfont\ssfam=\niness \scriptfont\ssfam=\sixss
                            \scriptscriptfont\ssfam=\sixss
                            \def\ss{\fam\ssfam\niness}%
    \normalbaselineskip=12pt
    \setbox\strutbox=\hbox{\vrule height8.0pt depth3.0pt width0pt}%
    \let\big=\ninebig
    \normalbaselines\rm}

\def\eightpoint{\def\rm{\fam0\eightrm}%
    \textfont0=\eightrm      \scriptfont0=\sixrm
                             \scriptscriptfont0=\fiverm
    \textfont1=\eighti       \scriptfont1=\sixi
                             \scriptscriptfont1=\fivei
    \textfont2=\eightsy      \scriptfont2=\sixsy
                             \scriptscriptfont2=\fivesy
    \textfont3=\tenex        \scriptfont3=\tenex
                             \scriptscriptfont3=\tenex
    \textfont\itfam=\eightit \def\it{\fam\itfam\eightit}%
    \textfont\slfam=\eightsl \def\sl{\fam\slfam\eightsl}%
    \textfont\bffam=\eightbf \scriptfont\bffam=\sixbf
                             \scriptscriptfont\bffam=\fivebf
                             \def\bf{\fam\bffam\eightbf}%
    \textfont\ssfam=\eightss \scriptfont\ssfam=\sixss
                             \scriptscriptfont\ssfam=\sixss
                             \def\ss{\fam\ssfam\eightss}%
    \normalbaselineskip=10pt
    \setbox\strutbox=\hbox{\vrule height7.0pt depth2.0pt width0pt}%
    \let\big=\eightbig
    \normalbaselines\rm}

\def\tenbig#1{{\hbox{$\left#1\vbox to8.5pt{}\right.\n@space$}}}
\def\ninebig#1{{\hbox{$\textfont0=\tenrm\textfont2=\tensy
                       \left#1\vbox to7.25pt{}\right.\n@space$}}}
\def\eightbig#1{{\hbox{$\textfont0=\ninerm\textfont2=\ninesy
                       \left#1\vbox to6.5pt{}\right.\n@space$}}}

\font\sectionfont=cmbx10
\font\subsectionfont=cmti10

\def\figurecaptionfont{\ninepoint}
\def\tablecaptionfont{\ninepoint}


\newcount\equationno
\newcount\bibitemno
\newcount\figureno
\newcount\tableno

\equationno=0
\bibitemno=0
\figureno=0
\tableno=0
\advance\pageno by -1


\footline={\ifnum\pageno=0{\hfil}\else
{\hss\rm\the\pageno\hss}\fi}


\def\section #1. #2 \par
{\vskip0pt plus .20\vsize\penalty-150 \vskip0pt plus-.20\vsize
\vskip 1.6 true cm plus 0.2 true cm minus 0.2 true cm
\global\def\equationlabel{#1}
\global\equationno=0
\centerline{\sectionfont #1. #2}\par
\immediate\write\terminal{Section #1. #2}
\vskip 0.7 true cm plus 0.1 true cm minus 0.1 true cm}


\def\subsection #1 \par
{\vskip0pt plus 0.6 true cm\penalty-50 \vskip0pt plus-0.6 true cm
\vskip2.5ex plus 0.1ex minus 0.1ex
\leftline{\subsectionfont #1}\par
\immediate\write\terminal{Subsection #1}
\vskip1.0ex plus 0.1ex minus 0.1ex
\noindent}


\def\appendix #1 \par
{\vskip0pt plus .20\vsize\penalty-150 \vskip0pt plus-.20\vsize
\vskip 1.6 true cm plus 0.2 true cm minus 0.2 true cm
\global\def\equationlabel{\hbox{\rm#1}}
\global\equationno=0
\centerline{\sectionfont Appendix #1}\par
\immediate\write\terminal{Appendix #1}
\vskip 0.7 true cm plus 0.1 true cm minus 0.1 true cm}


\def\enum{\global\advance\equationno by 1
(\equationlabel.\the\equationno)}


\def\ifundefined#1{\expandafter\ifx\csname#1\endcsname\relax}

\def\ref#1{\ifundefined{#1}?\immediate\write\terminal{unknown reference
on page \the\pageno}\else\csname#1\endcsname\fi}

\newwrite\terminal
\newwrite\bibitemlist

\def\bibitem#1#2\par{\global\advance\bibitemno by 1
\immediate\write\bibitemlist{\string\def
\expandafter\string\csname#1\endcsname
{\the\bibitemno}}
\item{[\the\bibitemno]}#2\par}

\def\beginbibliography{
\vskip0pt plus .20\vsize\penalty-150 \vskip0pt plus-.20\vsize
\vskip 1.6 true cm plus 0.2 true cm minus 0.2 true cm
\centerline{\sectionfont References}\par
\immediate\write\terminal{References}
\immediate\openout\bibitemlist=biblist.aux
\frenchspacing
\vskip 0.7 true cm plus 0.1 true cm minus 0.1 true cm}

\def\endbibliography{
\immediate\closeout\bibitemlist
\nonfrenchspacing}


\def\figurecaption#1{\global\advance\figureno by 1
\narrower\figurecaptionfont
Fig.~\the\figureno. #1}

\def\tablecaption#1{\global\advance\tableno by 1
\vbox to 0.5 true cm { }
\centerline{\tablecaptionfont%
Table~\the\tableno. #1}
\vskip-0.4 true cm}

\tenpoint

\immediate\openin\bibitemlist=biblist.aux
\ifeof\bibitemlist\immediate\closein\bibitemlist
\else\immediate\closein\bibitemlist
\input biblist.aux \fi


\def\blackboardrrm{\mathchoice
{\rm I\kern-0.21 em{R}}{\rm I\kern-0.21 em{R}}
{\rm I\kern-0.19 em{R}}{\rm I\kern-0.19 em{R}}}

\def\blackboardzrm{\mathchoice
{\rm Z\kern-0.32 em{Z}}{\rm Z\kern-0.32 em{Z}}
{\rm Z\kern-0.28 em{Z}}{\rm Z\kern-0.28 em{Z}}}

\def\blackboardh{\mathchoice
{\ss I\kern-0.14 em{H}}{\ss I\kern-0.14 em{H}}
{\ss I\kern-0.11 em{H}}{\ss I\kern-0.11 em{H}}}

\def\blackboardp{\mathchoice
{\ss I\kern-0.14 em{P}}{\ss I\kern-0.14 em{P}}
{\ss I\kern-0.11 em{P}}{\ss I\kern-0.11 em{P}}}

\def\blackboardt{\mathchoice
{\ss T\kern-0.52 em{T}}{\ss T\kern-0.52 em{T}}
{\ss T\kern-0.40 em{T}}{\ss T\kern-0.40 em{T}}}

\def\thicktablerule{\hrule height1pt}
\def\thintablerule{\hrule height0.4pt}


\def\thisyear{\number\year}

\def\thismonth{\ifcase\month\or
January\or February\or March\or April\or May\or June\or
July\or August\or September\or October\or November\or December\fi}

\input epsf

%


\def\rmd{{\rm d}}

\def\rmO{{\rm O}}


\def\Re{{\rm Re}\,}


\def\proof{\noindent{\sl Proof:}\kern0.6em}

\def\frac#1#2{\hbox{$#1\over#2$}}
\def\dual{\mathstrut^*\kern-0.1em}

\def\lvec#1{\setbox0=\hbox{$#1$}
    \setbox1=\hbox{$\scriptstyle\leftarrow$}
    #1\kern-\wd0\smash{
    \raise\ht0\hbox{$\raise1pt\hbox{$\scriptstyle\leftarrow$}$}}
    \kern-\wd1\kern\wd0}
\def\rvec#1{\setbox0=\hbox{$#1$}
    \setbox1=\hbox{$\scriptstyle\rightarrow$}
    #1\kern-\wd0\smash{
    \raise\ht0\hbox{$\raise1pt\hbox{$\scriptstyle\rightarrow$}$}}
    \kern-\wd1\kern\wd0}


\def\nabstar#1{\nabla\kern-0.5pt\smash{\raise 4.5pt\hbox{$\ast$}}
               \kern-4.5pt_{#1}}
\def\drv#1{{\partial_{#1}}}
\def\drvstar#1{\partial\kern-0.5pt\smash{\raise 4.5pt\hbox{$\ast$}}
               \kern-5.0pt_{#1}}
\def\drvtilde#1#2{{\tilde{\partial}_{#1}^{#2}}}


\def\momp#1#2{
    \setbox0=\hbox{${#1}$}\setbox1=\hbox{${#1}_{#2}$}
    {#1}_{#2}\kern-\wd1\kern\wd0
    \smash{\raise4.5pt\hbox{$\scriptscriptstyle +$}}}
\def\momm#1#2{
    \setbox0=\hbox{${#1}$}\setbox1=\hbox{${#1}_{#2}$}
    {#1}_{#2}\kern-\wd1\kern\wd0
    \smash{\raise4.5pt\hbox{$\scriptscriptstyle -$}}}
\def\mompm#1#2{
    \setbox0=\hbox{${#1}$}\setbox1=\hbox{${#1}_{#2}$}
    {#1}_{#2}\kern-\wd1\kern\wd0
    \smash{\raise4.5pt\hbox{$\scriptscriptstyle \pm$}}}
\def\smomp#1#2{
    \setbox0=\hbox{${#1}$}\setbox1=\hbox{${#1}_{#2}$}
    {#1}_{#2}\kern-\wd1\kern\wd0
    \smash{\raise3pt\hbox{$\scriptscriptstyle +$}}}
\def\smomm#1#2{
    \setbox0=\hbox{${#1}$}\setbox1=\hbox{${#1}_{#2}$}
    {#1}_{#2}\kern-\wd1\kern\wd0
    \smash{\raise3pt\hbox{$\scriptscriptstyle -$}}}
\def\smompm#1#2{
    \setbox0=\hbox{${#1}$}\setbox1=\hbox{${#1}_{#2}$}
    {#1}_{#2}\kern-\wd1\kern\wd0
    \smash{\raise3pt\hbox{$\scriptscriptstyle \pm$}}}




\def\psibar{\overline{\psi}}

\def\rhoprime{\rho\kern1pt'}
\def\rhobar{\bar{\rho}}
\def\rhobarprime{\rhobar\kern1pt'}
\def\rhobartilde{\kern2pt\tilde{\kern-2pt\rhobar}}
\def\rhobartildeprime{\kern2pt\tilde{\kern-2pt\rhobar}\kern1pt'}

\def\zetabar{\bar{\zeta}}
\def\zetaprime{\zeta\kern1pt'}
\def\zetabarprime{\zetabar\kern1pt'}
\def\zetar{\zeta_{\raise-1pt\hbox{\sixrm R}}}
\def\zetabarr{\zetabar_{\raise-1pt\hbox{\sixrm R}}}

\def\phiimpr{\phi_{\kern0.5pt\hbox{\sixrm I}}}

\def\ar{A_{\hbox{\sixrm R}}}

\def\vr{V_{\hbox{\sixrm R}}}
\def\aimpr{A_{\hbox{\sixrm I}}}

\def\vimpr{V_{\hbox{\sixrm I}}}
\def\op{{\cal O}}
\def\opprime{{{\cal O}^{\kern1pt\smash{\hbox{$\scriptstyle\prime$}}}}}


\def\dirac#1{\gamma_{#1}}
\def\diracstar#1#2{
    \setbox0=\hbox{$\gamma$}\setbox1=\hbox{$\gamma_{#1}$}
    \gamma_{#1}\kern-\wd1\kern\wd0
    \smash{\raise4.5pt\hbox{$\scriptstyle#2$}}}


\def\ba{b_{\rm A}}
\def\bv{b_{\rm V}}

\def\bg{b_{\rm g}}

\def\bx{b_{\rm X}}

\def\ca{c_{\rm A}}

\def\csw{c_{\rm sw}}

\def\cv{c_{\rm V}}


\def\faprime{\setbox0=\hbox{$f$}\setbox1=\hbox{$f_{\rm A}$}
    f_{\rm A}\kern-\wd1\kern\wd0
    \smash{\raise5.5pt\hbox{\kern0.5pt$\scriptstyle\prime$}}\kern1pt}
\def\fpprime{\setbox0=\hbox{$f$}\setbox1=\hbox{$f_{\rm P}$}
    f_{\rm P}\kern-\wd1\kern\wd0
    \smash{\raise5.5pt\hbox{\kern0.5pt$\scriptstyle\prime$}}\kern1pt}
\def\fxprime{\setbox0=\hbox{$f$}\setbox1=\hbox{$f_{\rm X}$}
    f_{\rm X}\kern-\wd1\kern\wd0
    \smash{\raise5.5pt\hbox{\kern0.5pt$\scriptstyle\prime$}}\kern1pt}
\def\fv{f_{\rm V}}
\def\fvi{f_{\rm V}^{{\kern1pt\hbox{\sixrm I}}}}
\def\fvr{f_{\rm V}^{{\kern1pt\hbox{\sixrm R}}}}
\def\faai{f_{\rm AA}^{{\kern1pt\hbox{\sixrm I}}}}
\def\fai{f_{\rm A}^{{\kern1pt\hbox{\sixrm I}}}}
\def\faa{f_{\rm AA}}
\def\fpa{f_{\rm PA}}
\def\fap{f_{\rm AP}}
\def\fpp{f_{\rm PP}}


\def\da{\delta_{\hbox{\sixrm A}}}
\def\dv{\delta_{\hbox{\sixrm V}}}


\def\tr{\,\hbox{tr}\,}



\def\gparisi{g_{{\hbox{\sixrm P}}}}

\def\mq{m_{\rm q}}

\def\mav{m_{\rm av}}

\def\za{Z_{\rm A}}

\def\zv{Z_{\rm V}}

\def\gtilde{\tilde{g}_0}

\def\hop{\kappa}
\def\hopc{\kappa_{\rm c}}

\def\msbar{{\rm \overline{MS\kern-0.05em}\kern0.05em}}

\rightline{OUTP 96-64P}
\rightline{CERN-TH/96-312}
\rightline{DESY 96-222}
\rightline{FSU-SCRI-96-116}

\vskip 1.0 true cm minus 0.3 true cm
\centerline
{\bigbf Non-perturbative determination of the axial current} 
\vskip2ex
\centerline
{\bigbf normalization constant in O(a) improved lattice QCD}
\vskip 1.2 true cm
\leftline{ 
Martin L\"{u}scher$^{\rm a}$,
Stefan Sint$^{\rm b}$,
Rainer Sommer$^{\rm c,d}$ and
Hartmut Wittig$^{\rm e}$}
\vskip 0.5 true cm
\parindent=1em

\item{$^{\rm a}$} 
Deutsches Elektronen-Synchrotron DESY, \hfill\break
Notkestrasse 85, D-22603 Hamburg, Germany

\vskip1ex
\item{$^{\rm b}$} 
SCRI, The Florida State University, Tallahassee, FL 32306-4052, USA

\vskip1ex
\item{$^{\rm c}$}
CERN, Theory Division, CH-1211 Gen\`eve 23,
Switzerland

\vskip1ex
\item{$^{\rm d}$}
DESY-IfH Zeuthen, Platanenallee 6, D-15738 Zeuthen, Germany

\vskip1ex
\item{$^{\rm e}$}
Theoretical Physics, University of Oxford,\hfill\break
1 Keble Road, Oxford OX1 3NP, England

\parindent=2em
\vskip 2.0 true cm
\centerline{\bf Abstract}
\vskip 1.5ex
A finite-size technique
is employed to compute the normalization constant $\za$
of the isovector axial current in lattice QCD.
The calculation is carried out in the quenched approximation for 
values of the bare gauge coupling $g_0$ ranging from $0$ to $1$.
In the lattice action and the lattice expression for the axial current
we include the counterterms required for O($a$) improvement,
with non-perturbatively determined coefficients. 
With little additional work 
the normalization constant $\zv$ of 
the improved isospin current is also obtained.
\vfill
\centerline{\thismonth\space\thisyear}
\eject

\section 1. Introduction

In lattice QCD with Wilson quarks 
the conservation of the isovector axial current is violated by 
lattice effects. As a consequence 
a finite renorma\-li\-zation of the current is required 
to ensure that the chiral Ward identities assume their
canonical form [\ref{BochicchioEtAl},\ref{MaianiMartinelli}].
It is evidently important to compute 
the associated normalization constant $\za$,
since it contributes directly to physical matrix elements such as
the pion decay constant $F_{\pi}$.

In perturbation theory $\za$
has been worked out to one-loop order
for various lattice actions and
lattice definitions of the axial current 
[\ref{MeyerSmith}--\ref{BorelliEtAl}].
The results may be used to calculate $\za$
at the couplings of interest, but since these are not small in general
it is difficult to say how reliable the numbers are 
that one obtains. A non-perturbative determination of 
the normalization constant is clearly preferable.
Two different strategies to perform such a calculation
have been pursued. In the first case $\za$ is fixed
by requiring certain chiral Ward identities between
correlation functions of the axial and vector currents to be satisfied
on the lattice
[\ref{MaianiMartinelli},\ref{MartinelliEtAlI}--\ref{HentyEtAl}].
The correlation functions are then evaluated through numerical simulation.
The other proposition is to 
compute matrix elements of the axial 
current between quark states and to 
determine the normalization of the current by 
matching the numerical results with renormalized perturbation 
theory at large momentum transfers [\ref{MartinelliEtAlII}].

Our principal aim in the present paper is to calculate $\za$
in the on-shell O($a$) improved lattice theory.
The significance of improvement in this context has previously 
been stressed in refs.~[\ref{HeatlieEtAl}--\ref{MartinelliEtAlIV}].
Here we employ the improved action and the improved axial current with 
non-perturbatively determined O($a$) counterterms
[\ref{letter}--\ref{paperIII}].
All calculations are carried out in the quenched approximation.
We use the Ward identity method and combine it with 
a finite-size technique based on the Schr\"odinger functional.
This allows us to set the quark mass to zero 
(or to values very close to zero)
and to determine $\za$ at all bare couplings $g_0$ between $0$ and $1$.
Contact with perturbation theory can thus be made.

For the definition of the Schr\"odinger functional
and the O($a$) improved theory the reader is referred to 
ref.~[\ref{paperI}].
The notations introduced there are taken over completely without
further notice.
In sect.~2 we briefly recall the euclidean Ward identities
associated with the chiral symmetry of QCD in the continuum limit.
We then define the 
isospin vector and axial vector currents in the on-shell O($a$)
improved lattice theory and derive
the normalization conditions that will be used to 
compute the associated normalization constants
(sect.~3).
As discussed in sect.~4 
a careful interpretation of the r\^ole played by the current 
normalization conditions is required on the lattice,
because the chiral Ward identities 
are only valid up to cutoff effects of order $a^2$.
The calculation of the isospin
vector and axial vector current normalization constants
through numerical simulations 
is described in sects.~5 and 6.
The paper ends with a few concluding remarks and 
a technical appendix, where an essentially rigorous
proof of the crucial Ward identity is given.

\section 2. Euclidean current algebra 

We first consider the theory in the continuum limit and proceed
formally, i.e.~without paying attention to the proper definition
of the correlation functions that occur.
The boundary conditions on the quark and gluon fields 
do not matter in this section.
We assume that there is an isospin doublet of
quarks with mass $m$ and study the 
associated chiral symmetry of the theory.

The isospin vector and axial vector variations of the 
quark and anti-quark field are defined by
$$
  \eqalignno{
  \dv^a\psi(x)&=\frac{1}{2}\tau^a\psi(x),
  \qquad\quad\quad\!
  \dv^a\psibar(x)=-\psibar(x)\frac{1}{2}\tau^a,
  &\enum\cr
  \noalign{\vskip2ex}
  \da^a\psi(x)&=\frac{1}{2}\tau^a\dirac{5}\psi(x),
  \qquad\phantom{\dirac{5}}
  \da^a\psibar(x)=\psibar(x)\dirac{5}\frac{1}{2}\tau^a,
  &\enum\cr}
$$
where $\tau^a$ denotes a Pauli matrix acting on the flavour indices of the
quark field. The definition extends to 
arbitrary expressions $\cal O$ by treating $\dv^a$
and $\da^a$ as first order differential operators.
In particular, for the variations of the isospin vector and 
axial vector currents,
$$
  V^a_{\mu}(x)=
  \psibar(x)\dirac{\mu}\frac{1}{2}\tau^a\psi(x),
  \qquad\quad
  A^a_{\mu}(x)=
  \psibar(x)\dirac{\mu}\dirac{5}\frac{1}{2}\tau^a\psi(x),
  \eqno\enum
$$
one obtains
$$
  \eqalignno{
  \dv^aV^b_{\mu}(x)&=-i\epsilon^{abc}V^c_{\mu}(x),
  \qquad\,
  \da^aV^b_{\mu}(x)=-i\epsilon^{abc}A^c_{\mu}(x),
  &\enum\cr
  \noalign{\vskip2ex}
  \dv^aA^b_{\mu}(x)&=-i\epsilon^{abc}A^c_{\mu}(x),
  \qquad
  \da^aA^b_{\mu}(x)=-i\epsilon^{abc}V^c_{\mu}(x).
  &\enum\cr}
$$
The currents thus form a closed algebra under these variations.

The Ward identities associated with
the chiral symmetry of the action
are derived by performing local infinitesimal symmetry transformations
of the quark and anti-quark fields in the euclidean functional integral.
We choose to write the identities in an integrated
form which is quite intuitive and will prove useful later on
when we discuss the lattice theory.

Let $R$ be a space-time region with smooth boundary $\partial R$.
Suppose ${\cal O}_{\rm int}$ and ${\cal O}_{\rm ext}$ 
are polynomials in the basic fields localized
in the interior and exterior of $R$ respectively.
The general vector current Ward identity then reads
$$
  \int_{\partial R}\rmd\sigma_{\mu}(x)\,  
  \left\langle 
  V^a_{\mu}(x) {\cal O}_{\rm int} {\cal O}_{\rm ext} 
  \right\rangle
  =-
  \left\langle 
  \left(\dv^a{\cal O}_{\rm int}\right) {\cal O}_{\rm ext} 
  \right\rangle,
  \eqno\enum
$$
while for the axial current one obtains
$$
  \eqalignno{
  \int_{\partial R}\rmd\sigma_{\mu}(x)\,  
  \left\langle 
  A^a_{\mu}(x) {\cal O}_{\rm int} {\cal O}_{\rm ext} 
  \right\rangle
  &=-
  \left\langle 
  \left(\da^a{\cal O}_{\rm int}\right) {\cal O}_{\rm ext} 
  \right\rangle
  &\cr
  \noalign{\vskip2ex}
  &+
  2m\int_R\rmd^4x\,
  \left\langle 
  P^a(x) {\cal O}_{\rm int} {\cal O}_{\rm ext} 
  \right\rangle.
  &\enum\cr}
$$
The integration measure $\rmd\sigma_{\mu}(x)$ points 
along the outward normal to the surface $\partial R$
and the pseudo-scalar density $P^a(x)$ is defined by
$$
  P^a(x)=\psibar(x)\dirac{5}\frac{1}{2}\tau^a\psi(x).
  \eqno\enum
$$
The left-hand sides of eqs.~(2.6) and (2.7) may be interpreted as
matrix elements of the charge operators associated with the currents.
This is made particularly clear if we choose
$R$ to be the space-time volume between two equal-time hyper-planes.

For illustration we set the quark mass to zero
and choose ${\cal O}_{\rm int}$ to be one of the currents at some point $y$ in
the interior of $R$. 
An example of the resulting identities then is 
$$
  \int_{\partial R}\rmd\sigma_{\mu}(x)\, 
  \left\langle 
  A^a_{\mu}(x) A^b_{\nu}(y) {\cal O}_{\rm ext} 
  \right\rangle
  =
  i\epsilon^{abc}
  \left\langle 
  V^c_{\nu}(y){\cal O}_{\rm ext} 
  \right\rangle,
  \eqno\enum
$$
and three more such relations, corresponding to eqs.~(2.4) and (2.5),
may be obtained.
In this way the current algebra, which one usually sets up in Minkowski space
in an operator language, is recovered in the euclidean domain.

\section 3. Currents in lattice QCD

As already mentioned in sect.~1 
we choose Wilson's formulation of lattice QCD
and include all O($a$) correction terms that are required for
on-shell improvement. The lattice action etc.~is exactly as 
in ref.~[\ref{paperI}]. We assume
that the coefficients multiplying the O($a$) counterterms have
been adjusted so that the residual cutoff effects are of order $a^2$.

\subsection 3.1 Improved currents

On the lattice the bare currents,
$V_{\mu}^a$ and $A_{\mu}^a$,
are defined through the local expressions (2.3).
The unrenormalized on-shell O($a$) improved currents are then given by 
[\ref{letter},\ref{paperI}]
$$
  \eqalignno{
  (\vimpr)_{\mu}^a&=V_{\mu}^a
  +\cv a\frac{1}{2}(\drvstar{\nu}+\drv{\nu})T_{\mu\nu}^a,
  &\enum\cr
  \noalign{\vskip2ex}
  (\aimpr)_{\mu}^a&=A_{\mu}^a
  +\ca a\frac{1}{2}(\drvstar{\mu}+\drv{\mu})P^a.
  &\enum\cr}
$$
For the anti-symmetric tensor field $T_{\mu\nu}^a$ we can take
$$
  T_{\mu\nu}^a(x)=i\psibar(x)\sigma_{\mu\nu}\frac{1}{2}\tau^a\psi(x),
  \eqno\enum
$$
while the axial density $P^a$ is again defined through eq.~(2.8).

The renormalization of lattice QCD is particularly transparent
if a mass-independent renormalization scheme is employed.
As discussed in sect.~3 of ref.~[\ref{paperI}] the renormalized 
improved currents in such schemes are given by
$$
  \eqalignno{
  (\vr)_{\mu}^a&=\zv(1+\bv a\mq)(\vimpr)_{\mu}^a,
  &\enum\cr
  \noalign{\vskip2ex}
  (\ar)_{\mu}^a&=\za(1+\ba a\mq)(\aimpr)_{\mu}^a.
  &\enum\cr}
$$
The renormalization constants $\zv$ and $\za$ are functions of
the modified bare coupling
$$
  \gtilde^2=g_0^2(1+\bg a\mq),
  \eqno\enum
$$
while $\bv$, $\ba$ and $\bg$ depend on $g_0$ and should be adjusted 
so as to cancel any remaining cutoff effects of order $a\mq$.
We shall not need to know these coefficients here,
because the quark mass will be set to zero 
for the calculation of $\zv$ and $\za$.

\subsection 3.2 Normalization condition for the vector current

Although the isospin symmetry of the continuum theory is preserved on 
the lattice, the improved vector current introduced above 
is only conserved up to cutoff effects of order $a^2$.
Its normalization is hence not naturally given and we 
must impose a normalization condition to fix $\zv$.
Our aim in the following is to derive such a condition 
by studying the action of the renormalized isospin charge 
on states with definite isospin quantum numbers.

The matrix elements that we shall consider are constructed in the 
framework of the Schr\"odinger functional (see ref.~[\ref{paperI}]
for details). We use the boundary field products
$$
  \eqalignno{
  \op^a&=a^6\sum_{\bf u,v}
  \zetabar({\bf u})\dirac{5}\frac{1}{2}
  \tau^a\zeta({\bf v}),
  &\enum\cr
  \noalign{\vskip2ex}
  \opprime^a&=a^6\sum_{\bf u,v}
  \zetabarprime({\bf u})\dirac{5}\frac{1}{2}
  \tau^a\zetaprime({\bf v}),
  &\enum\cr}
$$
to create initial and final states that transform according  
to the vector representation of the exact isospin symmetry.
The correlation function 
$$
  \fvr(x_0)={a^3\over6L^6}\sum_{\bf x}i\epsilon^{abc}
  \langle\opprime^a(\vr)_0^b(x)\op^c\rangle
  \eqno\enum
$$
can then be interpreted as a matrix element of the renormalized isospin
charge between such states. The charge generates an infinitesimal
isospin rotation (if properly normalized) and 
after some algebra one finds that the correlation function must be equal to
$$
  f_1=-{1\over3L^6}
  \langle\opprime^a\op^a\rangle.
  \eqno\enum
$$
Strictly speaking this argumentation is only correct in the continuum theory.
We may however conclude that 
$$
  \fvr(x_0)=f_1+\rmO(a^2),
  \eqno\enum
$$
since the lattice correlation functions approach the continuum limit
with a rate proportional to $a^2$.
Note that we do not need to include 
the renormalization factors for the boundary quark fields here
because they cancel in eq.~(3.11).

The O($a$) counterterm appearing in the definition (3.1) of the improved
vector current does not contribute to the correlation function
$\fvr(x_0)$. So if we introduce the analogous correlation function
for the bare current,
$$
  \fv(x_0)={a^3\over6L^6}\sum_{\bf x}i\epsilon^{abc}
  \langle\opprime^aV_0^b(x)\op^c\rangle,
  \eqno\enum
$$
it follows from eq.~(3.11) that 
$$
  \zv(1+\bv a\mq)\fv(x_0)=f_1+\rmO(a^2).
  \eqno\enum
$$
By evaluating the correlation functions $f_1$ and $\fv(x_0)$
through numerical simulation
one is thus able to compute the normalization factor
$\zv(1+\bv a\mq)$.
In particular, to calculate $\zv$ it suffices to consider the 
theory at vanishing quark mass.

\subsection 3.3 Normalization condition for the axial current

To derive a normalization condition for $\za$,
we set the quark mass to zero from 
the beginning. Our starting point is the Ward identity 
(2.9) which we now write in the form
$$
  \int_{\partial R}\rmd\sigma_{\mu}(x)\, 
  \epsilon^{abc}
  \left\langle 
  A^a_{\mu}(x) A^b_{\nu}(y) {\cal O}_{\rm ext} 
  \right\rangle
  =
  2i
  \left\langle 
  V^c_{\nu}(y){\cal O}_{\rm ext} 
  \right\rangle.
  \eqno\enum
$$
One may be hesitant to make use of this relation,
since it has been deduced in a formal manner.
A general argument, presented in appendix A,
however shows that such worries are not justified.
There is little doubt that eq.~(3.14) is a true property of 
the theory in the continuum limit and so may be used
to fix the normalization of the axial current on the lattice.

We now pass to the lattice theory and assume Schr\"odinger functional
boundary conditions as before. A convenient choice of the region
$R$ is the space-time volume between the hyper-planes at
$x_0=y_0\pm t$. From eq.~(3.14) and O($a$) improvement we then expect that
$$
  \eqalignno{
  &a^3\sum_{\bf x}
  \epsilon^{abc}
  \left\langle 
  [(\ar)^a_{0}(y_0+t,{\bf x})-(\ar)^a_{0}(y_0-t,{\bf x})]
  (\ar)^b_{0}(y) {\cal O}_{\rm ext} 
  \right\rangle
  &\cr
  \noalign{\vskip2ex}
  &\qquad\qquad\qquad =2i
  \left\langle 
  (\vr)^c_{0}(y){\cal O}_{\rm ext} 
  \right\rangle
  +\rmO(a^2).
  &\enum\cr}
$$
It has been important here that the fields 
in the correlation functions
are localized at non-zero distances from each other.
Since the theory is only on-shell improved, one would otherwise
not be able to say that the error term is of order $a^2$
(cf.~sect.~2 of ref.~[\ref{paperI}]).

\topinsert
\vbox{
\vskip0.0true cm

\centerline{
\epsfxsize=5.5 true cm
\epsfbox{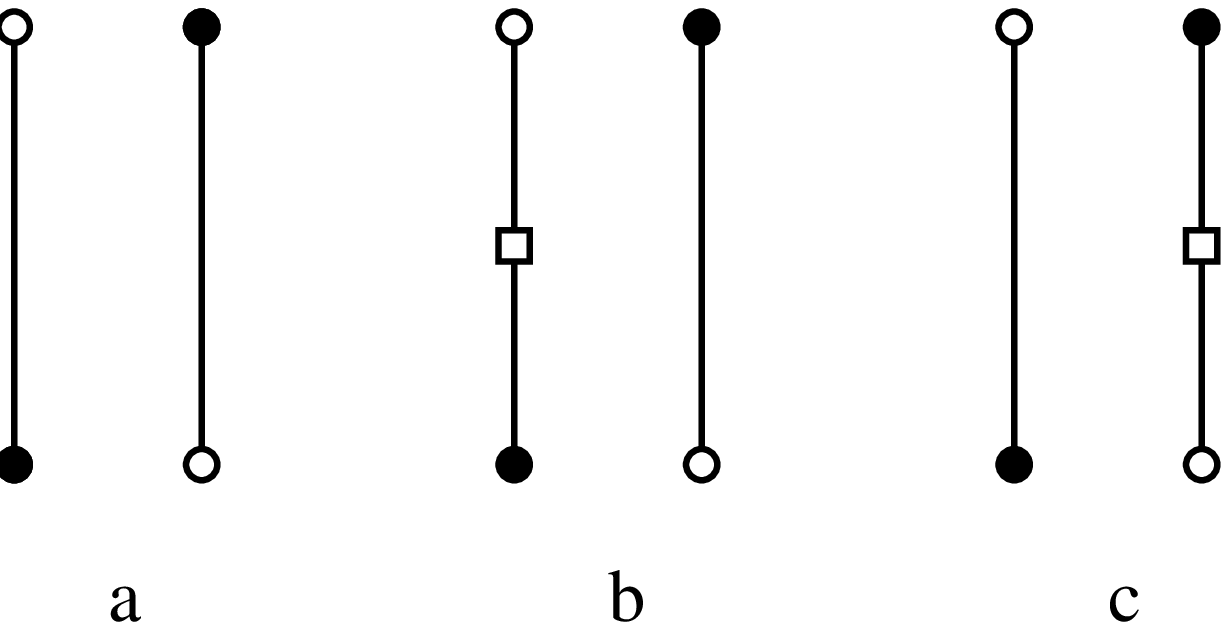}}

\vskip0.6true cm
\figurecaption{
Quark diagrams contributing to $f_1$ (diagram a) and 
$\fv(x_0)$ (diagrams b and c).
Filled (open) circles represent the creation (annihilation)
of a quark at the boundaries of the lattice.
The squares indicate the vector current insertions.
}
}
\endinsert

After summing over the spatial components of $y$,
and using the fact that the axial
charge is conserved at zero quark mass (up to corrections of order $a^2$),
eq.~(3.15) becomes
$$
  \eqalignno{
  &a^6\sum_{\bf x,y}
  \epsilon^{abc}
  \left\langle 
  (\ar)^a_{0}(x)
  (\ar)^b_{0}(y) {\cal O}_{\rm ext} 
  \right\rangle
  =
  &\cr
  \noalign{\vskip2ex}
  &\kern7em a^3\sum_{\bf y}i
  \left\langle 
  (\vr)^c_{0}(y){\cal O}_{\rm ext} 
  \right\rangle
  +\rmO(a^2),
  &\enum\cr}
$$
where $x_0=y_0+t$.
We now choose the field product ${\cal O}_{\rm ext}$ so that 
the function $\fvr(y_0)$ introduced previously appears on the 
right-hand side of eq.~(3.16).
The normalization condition for the vector current then allows us
to replace the correlation function by $f_1$.
Explicitly, we define
$$
  \faai(x_0,y_0)=-{a^6\over6L^6}\sum_{\bf x,y}
  \epsilon^{abc}\epsilon^{cde}
  \langle\opprime^d(\aimpr)_0^a(x)(\aimpr)_0^b(y)\op^e\rangle
  \eqno\enum
$$
and conclude from the above that 
$$
  \za^2\faai(x_0,y_0)=f_1+\rmO(a^2) 
  \eqno\enum
$$
for all times $x_0>y_0$ between $0$ and $T$.
The normalization constant $\za$ can thus be determined
by computing the correlation functions $f_1$ and $\faai(x_0,y_0)$
at zero quark mass.

\subsection 3.4 Disconnected diagrams and the strange quark

As usual the integration over the quark and anti-quark fields
in the functional integral is carried out analytically.
One is then left with an integration over all gauge fields,
the correlation functions $f_1$ etc.~being 
given by a set of quark diagrams 
that correspond to the possible Wick contractions 
of the quark and anti-quark fields. 
For Schr\"odinger functional boundary conditions
the required two-point contractions have been worked out
in detail in sect.~2 of ref.~[\ref{paperII}].

In the case of $f_1$ and $\fv(x_0)$ 
the assignment of the isospin quantum numbers
is such that no disconnected quark diagrams appear (see fig.~1).
The correlation function $\faai(x_0,y_0)$
involves two current insertions and many more Wick contractions
exist. Among the diagrams listed in fig.~2 there are 
also two disconnected ones (diagrams g and h).
The isospin factors associated with the diagrams e and f 
vanish so that they can be dropped immediately.

\topinsert
\vbox{
\vskip0.0true cm

\centerline{
\epsfxsize=8 true cm
\epsfbox{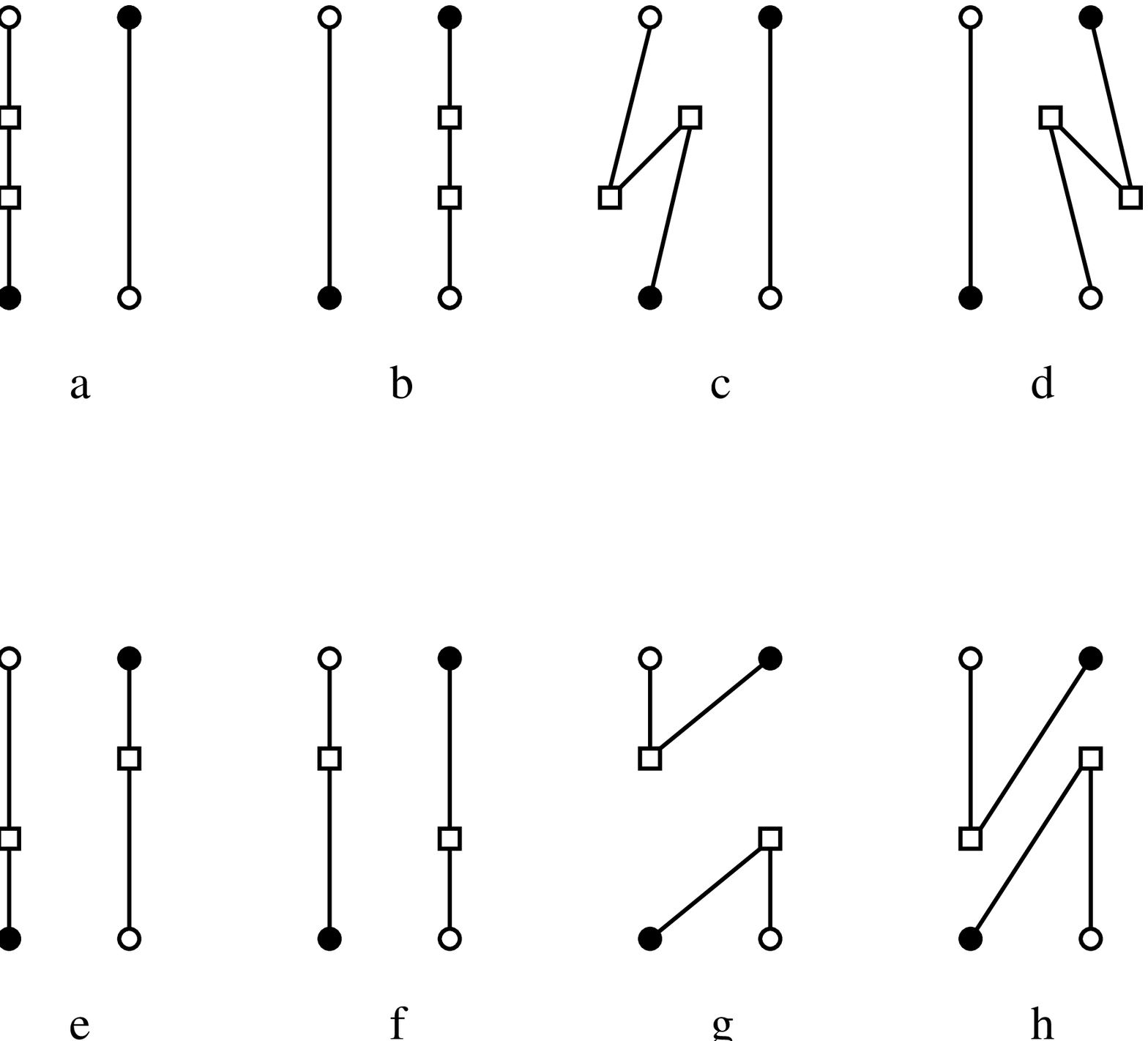}}

\vskip0.6true cm
\figurecaption{Quark diagrams contributing to $\faai(x_0,y_0)$.
The squares indicate the axial current insertions
at time $x_0$ (upper square) and $y_0$ (lower square).
}
}
\endinsert

We would now like to show that the disconnected diagrams do not contribute
either. To this end we 
introduce a third quark, referred to as the strange quark,
and replace the field products 
$\opprime^a$ and $\op^a$ in the definitions of 
$f_1$ and $\faai(x_0,y_0)$
by products of a strange and an isospin doublet quark boundary field.
With appropriately contracted indices, the argumentation 
leading to eq.~(3.18) then goes through unchanged.
Note that the currents are the same as before and 
the calculated value of $\za$ must 
hence come out to be the same up to terms of order $a^2$.
Since the strange quark does not couple to the axial current,
the number of possible Wick contractions is strongly reduced and   
only the diagrams a and c survive, the line without current insertions
representing the strange quark propagator.
If we interchange quarks and anti-quarks one obtains the diagrams
b and d instead. 
The isospin factors associated with the diagrams 
can be worked out straightforwardly and a comparison of 
the situation with and without strange quark then shows
that the total contribution of the disconnected 
diagrams must vanish up to terms of order $a^2$.

In the course of the numerical computations described later in 
this paper, we have been able to verify that 
the disconnected diagrams indeed add up to zero within small
statistical errors. We have thus decided 
to drop them and to extract
$\za$ from the connected part of $\faai(x_0,y_0)$.

\section 4. Lattice effects and current normalization 

The relations (3.13) and (3.18) determine the current normalization
constants $\zv$ and $\za$ only up to cutoff effects of order $a^2$.
Depending on the choice of the 
lattice size, the boundary values of the gauge field and the other
kinematical parameters that one has, slightly different results 
for $\zv$ and $\za$ are hence obtained. 
One may try to assign a 
systematic error to the normalization constants
by studying these variations in detail, but since there is no
general rule as to which choices of the kinematical parameters are 
considered to be reasonable, such error estimates are bound to be rather
subjective.

In our opinion the better way to deal
with the problem is to {\it define}\/ the normalization constants 
through a particular normalization condition.
The physical matrix elements of the renormalized currents 
that one is interested in must then be calculated 
for a range of lattice spacings so as 
to be able to extrapolate the data to the continuum limit.
The results obtained in this way are guaranteed to be independent
of the chosen normalization condition, because any differences 
in the normalization constants of order $a^2$ 
extrapolate to zero together with the 
cutoff effects associated with the matrix elements themselves.

In the following we shall adopt this point of view and the
precise choices that we shall make are then not too important.
Some care must be paid to ensure that  
the cutoff effects in matrix elements of the renormalized currents
between low-energy states are not artificially enhanced through 
an inappropriate choice of the 
kinematical parameters in eqs.~(3.13) and (3.18).
In particular, the external length scales 
(the time difference $x_0-y_0$ and the lattice size, for example)
should be sufficiently large compared to the lattice spacing,
at all bare couplings considered.
Perturbation theory can serve as a guide here
and further confidence can be gained by 
studying the magnitude of the residual cutoff effects in various
matrix elements of the renormalized currents through numerical simulations.

\section 5. Numerical evaluation of $f_1$, $\fv$ and $\faai$

In this section we present the details of the numerical calculations
which we have performed to determine $\za$ and $\zv$. We start in
subsect.~5.1 by listing the expressions for $f_1$, $\fv$ and $\faai$
in terms of quark propagators. 
In subsect.~5.2 we briefly discuss the
lattice action and algorithms used in the numerical simulation. We
have closely followed the procedures outlined in
ref.~[\ref{paperIII}], which can be consulted for further information
and any unexplained notations.

\subsection 5.1 Explicit form of $f_1$, $\fv$ and $\faai$

As already mentioned in subsect.~3.4,
the mathematical expressions corresponding to 
the quark diagrams shown in figs.~1 and 2 are obtained 
by applying Wick's theorem to the appropriate product of quark fields.
There is only one diagram contributing to
the correlation function $f_1$
and one finds that
$$
  f_1=
  \frac{1}{2}\left\langle\tr\{K^{\dagger}K\}\right\rangle_{\rm G},
  \eqno\enum
$$
where the trace is over the Dirac and colour indices. 
The matrix $K$
represents the quark propagation from the boundary 
at time~$0$ to the boundary at time~$T$.
In terms of the solution $H(x)$ of the lattice Dirac equation
introduced in sect.~2 of ref.~[\ref{paperIII}]
it is given by
$$
  K={a^3\over L^3}
  \sum_{\bf x}\left\{P_{+}U(x,0)^{-1}H(x)\right\}_{x_0=T-a}.
  \eqno\enum
$$
The numerical calculation of $H(x)$ is discussed in subsect.~3.2
of ref.~[\ref{paperIII}].

In the case of the correlation function $\fv(x_0)$,
two diagrams contribute and one obtains
$$
  \fv(x_0)=
  {a^3\over2L^3}\sum_{\bf x}\left\langle\Re\tr\{
  K^{\dagger}\dirac{5}H'(x)^{\dagger}\dirac{5}\dirac{0}H(x)
  \}\right\rangle_{\rm G},
  \eqno\enum
$$
where $H'(x)$ denotes the quark propagator from the
boundary at time~$T$ to the point~$x$ in the interior of the space-time
volume. $H'(x)$ is defined through the Dirac equation
$$
  (D+\delta D+m_0)H'(x)=0,\qquad
  0<x_0<T,
  \eqno\enum
$$
and the boundary conditions 
$$
  P_{+}H'(x)|_{x_0=0}=0,\qquad
  P_{-}H'(x)|_{x_0=T}=P_{-}.
  \eqno\enum
$$
The numerical solution of these equations proceeds
as in the analogous case of the matrix $H(x)$
(cf.~subsect.~3.2 of ref.~[\ref{paperIII}]).

Taking the definition (3.2) of the improved axial current into account,
the correlation function $\faai(x_0,y_0)$ may be expanded according to
$$
  \faai=\faa+
  \ca a\left[\drvtilde{0}{x}\fpa+\drvtilde{0}{y}\fap\right]+
  \ca^2 a^2\drvtilde{0}{x}\drvtilde{0}{y}\fpp
  \eqno\enum
$$
with the obvious definitions of $\faa$, $\fap$, $\fpa$ and $\fpp$.
The superscripts on the symmetric difference operator
$$
  \tilde{\partial}_\mu=\frac{1}{2}(\drvstar{\mu}+\drv{\mu})
  \eqno\enum
$$
imply a differentiation with respect to $x$ or $y$ respectively.
As discussed in subsect.~3.4 only the diagrams a-d in fig.~2 need to 
be evaluated for the computation of $\za$. The corresponding 
expressions for $\faa$, $\fap$, $\fpa$ and $\fpp$
look very similar and we only give the result for 
$\faa$ and diagram a, viz.
$$
  \eqalignno{
  &\left\{\faa(x_0,y_0)\right\}_{\hbox{\sevenrm diagram a}}=
  &\cr
  \noalign{\vskip2ex}
  &\qquad{a^6\over4L^3}\sum_{\bf x,y}
  \left\langle\tr\left\{
  K^{\dagger}\dirac{5}H'(x)^{\dagger}\dirac{5}\dirac{0}\dirac{5}
  S(x,y)\dirac{0}\dirac{5}H(y)
  \right\}\right\rangle_{\rm G}.
  &\enum\cr}
$$
The bulk quark propagator $S(x,y)$ appearing in this formula 
is the inverse of the Dirac operator $D+\delta D+m_0$ in the space
of quark fields with vanishing boundary values.
When evaluating $\faa$ and the other correlation functions, 
the propagator itself is however not required.
Instead one 
first calculates $H(x)$ at all $x$ and then solves 
the Dirac equation
$$
  (D+\delta D+m_0)
  \biggl\{a^3\sum_{\bf y}S(x,y)\dirac{0}\dirac{5}H(y)\biggr\}
  =a^{-1}\delta_{x_0y_0}\dirac{0}\dirac{5}H(x)
  \eqno\enum
$$
by applying the usual iterative methods.

\subsection 5.2 Details of the simulation

As for the coefficients multiplying the O($a$) counterterms
in the improved action and the improved axial current, 
we follow ref.~[\ref{paperIII}] and set
$$
  \eqalignno{
  \csw&={1-0.656\,g_0^2-0.152\,g_0^4-0.054\,g_0^6 \over
        1-0.922\,g_0^2},
  &\enum\cr
  \noalign{\vskip2ex}
  \ca&=-0.00756\,g_0^2\times{1-0.748\,g_0^2\over1-0.977\,g_0^2},
  &\enum\cr}
$$
for all couplings in the range $0\leq g_0\leq1$.
These formulae have been obtained non-perturbatively
by imposing some carefully chosen improvement conditions.
We expect that 
an almost perfect cancellation of O($a$) effects 
in on-shell quantities is thus achieved.

From now on we shall often quote values of $\beta=6/g_0^2$ and
$\kappa=(8+2am_0)^{-1}$ instead of the bare coupling and mass.
All our production runs have been performed on APE/Quadrics computers
with 256 nodes. We used the same hybrid over-relaxation algorithm as
described in subsect.~3.1 of ref.~[\ref{paperIII}] to generate a
representative ensemble of gauge field configurations. Subsequent
evaluations of the correlation functions $f_1$, $\fv$ and $\faai$ were
separated by~25 iterations of the algorithm. We have checked
explicitly for the statistical independence of our sample by dividing
the full ensemble into bins, each containing a number of individual
``measurements''. The statistical errors were then monitored as the
number of measurements per bin was increased. We did not observe any
significant change of the errors for increasing bin size, which we
take as evidence for the statistical independence of our sample.

To invert the Dirac operator we employed the
stabilized biconjugate gradient algorithm (BiCGstab) with even-odd
preconditioning~[\ref{BiCGstab},\ref{FrommerEtAl}]. By comparing our
results with those from a set of Fortran-90 programs, we have verified
not only the correct evaluation of correlation functions, but also
that the rounding errors associated with the 32~bit arithmetic on the
APE computer were completely negligible in our calculation.
All statistical errors were estimated using the jackknife
method.

\section 6. Computation of $\zv$ and $\za$

We now proceed to describe the non-perturbative determination of the
normalization factors $\zv$ and $\za$ in the range $0\leq{g_0}\leq1$. 
Except for the tests mentioned in subsect.~6.4, the boundary values 
$C$ and $C'$ of the gauge field and the angles $\theta_k$ are set to zero
throughout this section.

\subsection 6.1  Complete specification of the normalization conditions 

As discussed in sect.~4
we need to make a definite choice for 
the parameters on which
the normalization conditions (3.13) and (3.18) depend.
The quark mass is set to zero, as previously indicated,
and the remaining parameters are then   
the times at which the currents are inserted and
the lattice extensions $T$ and $L$.
We scale these parameters proportionally to $L$ and 
eventually decided to take
$$
  \eqalignno{
  &\zv\fv(T/2)=f_1,\qquad T=2L,
  &\enum\cr
  \noalign{\vskip2ex}
  &\za^2\faai(2T/3,T/3)=f_1,\qquad T=9L/4,
  &\enum\cr}
$$
as the definite form of the normalization conditions.
We still need to say, however, what precisely it means to set the quark mass
to zero and how $L/a$ is to be scaled with $g_0$.

The critical hopping parameter $\hopc$ (i.e.~the zero mass point)
depends on the details of the lattice definition of the quark mass
[\ref{paperI},\ref{paperIII}].
The ambiguity is just one
of the sources of the order $a^2$ corrections in the normalization
conditions for the currents
and so is to be treated following the lines of sect.~4, i.e.~we
adopt any particular definition of the 
quark mass and use it to calculate $\hopc$.

The definition that we have chosen is the same as the one 
previously employed in sect.~7 of ref.~[\ref{paperIII}]. 
The starting point is the 
unrenormalized current quark mass $m(x_0)$
which one extracts from the PCAC relation 
(eq.~(5.2) of ref.~[\ref{paperIII}]). 
We then set $T=2L$ and 
define $\hopc$ to be the value of the hopping parameter $\hop$ where 
$$
  \mav=\frac{1}{5}\sum_{t=-2a}^{2a}m(T/2+t)
  \eqno\enum
$$
vanishes. The average over the time coordinate $x_0$ is taken to reduce
the statistical error on the calculated mass values [\ref{paperIII}].
Note that the critical hopping parameter so defined is slightly dependent
on $L/a$. It is implicitly understood that at any given value of $\beta$
one chooses a lattice size $L/a$ and first computes $\hopc$
and then evaluates 
the normalization conditions (6.1) and (6.2) at this value of $\hopc$
and the same lattice size $L/a$.

At this point the normalization constants $\zv$ and $\za$ are 
well-defined functions of $g_0$ and $L/a$. 
The dependence on the lattice size is of order $(a/L)^2$ in the continuum limit
and we have verified that at the couplings of interest the change 
in the calculated values of the normalization constants is indeed small
when increasing $L/a$ from say $8$ to $16$. 

According to the 
discussion in sect.~4 we now need to make a definite choice of $L/a$.
To ensure that the on-shell matrix elements of the  
renormalized improved currents approach the continuum limit with a rate
proportional to $a^2$, we must require that $L$ remains fixed in 
physical units. Explicitly, we define $L/a$ at all couplings $g_0\leq1$ through
$$
  \eqalignno{
  L/a&=8\quad\hbox{at}\quad g_0=1,
  &\enum\cr
  \noalign{\vskip2ex}
  L/r_0&=\hbox{constant},
  &\enum\cr}
$$
where $r_0$ denotes a hadronic scale  
extracted from the force between heavy quarks [\ref{rnull}]. 
Using recent lattice data for $r_0$ [\ref{HartmutI}]
one finds that $L/a|_{\beta=6.2}\simeq11$, $L/a|_{\beta=6.4}\simeq14$
and $L/a>16$ for $\beta\geq6.8$.
For practical reasons we did not perform the simulations 
at exactly these lattice sizes. In fact this is not
really required 
because the associated systematic errors can be estimated reliably 
and turn out to be small
(details are given below).

\topinsert
\vbox{
\vskip-2.5true cm

\centerline{
\epsfxsize=11 true cm
\epsfbox{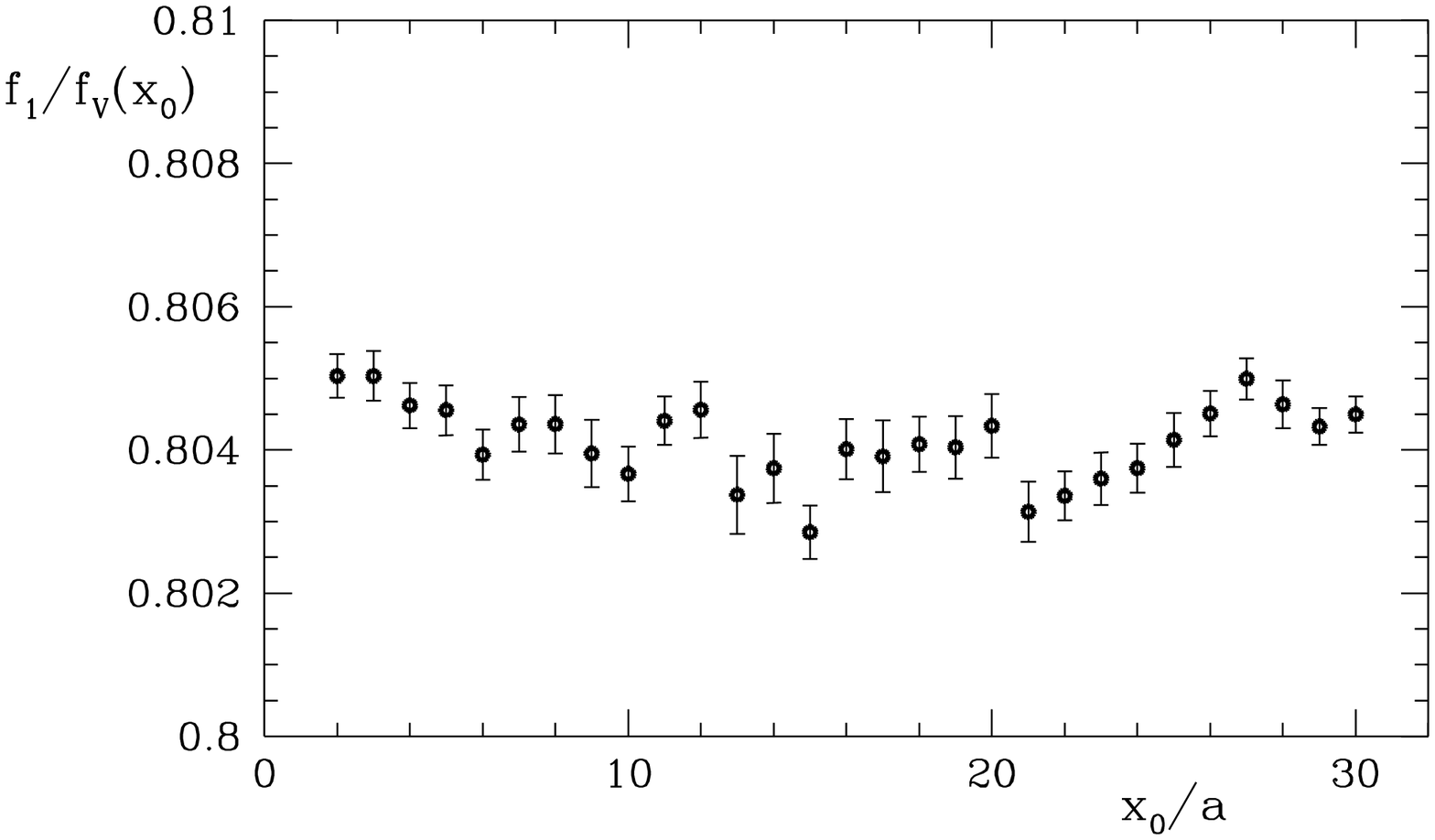}}

\vskip-1.5true cm
\figurecaption{Simulation results for $f_1/\fv(x_0)$  
from a lattice of size $32\times16^3$ at $\beta=6.4$
and zero quark mass. Note that we are using a fine scale in this plot.
The statistical fluctuations of the data points are on the level 
of a small fraction of a percent.
}
}
\endinsert

\subsection 6.2 Results for $\zv$ and $\za$

The statistical fluctuations of the numerators and denominators
in the ratios $f_1/\fv(x_0)$ and $f_1/\faai(x_0,y_0)$ are
strongly correlated. The jackknife error estimation 
accounts for these correlations and the ratios are obtained 
with impressive statistical accuracy even 
if only a small ensemble of 
independent gauge field configurations is available.
Fig.~3 shows the ratio $f_1/\fv(x_0)$ as a function of $x_0/a$ from a
typical run. 
One observes a clear signal with small statistical uncertainty
and nearly no time-dependence within errors.
The data points at different times $x_0$ are statistically decorrelated
to such an extent that the signal-to-noise ratio can be enhanced 
by averaging the data
in the range $T/2-2a\leq x_0\leq T/2+2a$.
Strictly speaking this should be taken as part of the definition 
of the normalization condition for $\zv$ 
(in the sense of sect.~4), but we did not want to obscure the discussion 
in subsect.~6.1 with too many details and thus mention this
item only now. In the case of the ratio $f_1/\faai(x_0,y_0)$
no averaging has been performed.

\vfill\eject

\topinsert
\newdimen\digitwidth
\setbox0=\hbox{\rm 0}
\digitwidth=\wd0
\catcode`@=\active
\def@{\kern\digitwidth}
\tablecaption{
The values of $L/a$ and $\hopc$ used in the normalization conditions
}
\vskip1ex
$$\vbox{\settabs\+x&xxxxx&%
                  xx&xxxx&%
                  xxxx&xxxxxxxxxxxx&
                  xxxxxxx&xxxx&%
                  xxxx&xxxxxxxxxxxx&\cr
\thicktablerule
\vskip1ex
                \+& \hfill     $\beta$ \kern0.3em
                 && \hfill     $L/a$   \kern0.5em
                 && \kern2.0em $\hopc$ \hfill
                 && \hfill     $L/a$   \kern0.5em
                 && \kern2.0em $\hopc$ \hfill
                 &  \cr
\vskip1.0ex
\thintablerule
\vskip1.5ex
  \+& \hfill $6.0$
  &&  \hfill $8$
  &&  $0.135046(16)$ \hfill
  &&\cr
  \+& \hfill $6.2$
  &&  \hfill $8$
  &&  $0.135692(6)$  \hfill
  &&  \hfill $12$
  &&  $0.135705(12)$ \hfill
  &&\cr
  \+& \hfill $6.4$
  &&  \hfill $8$
  &&  $0.135655(4)$  \hfill
  &&  \hfill $16$
  &&  $0.135720(9)$  \hfill
  &&\cr
  \+& \hfill $6.8$
  &&  \hfill $8$
  &&  $0.135078(8)$  \hfill
  &&  \hfill $16$
  &&  $0.135097(5)$  \hfill
  &&\cr
  \+& \hfill $7.4$
  &&  \hfill $8$
  &&  $0.134058(4)$  \hfill
  &&  \hfill $16$
  &&  $0.134071(4)$    \hfill
  &&\cr
  \+& \hfill $8.0$
  &&  \hfill $8$
  &&  $0.133168(4)$  \hfill
  &&  \hfill $16$
  &&  $0.133173(3)$    \hfill
  &&\cr
  \+& \hfill $9.6$
  &&  \hfill $8$
  &&  $0.131447(3)$  \hfill
  &&  \hfill $16$
  &&  $0.131448(2)$    \hfill
  &&\cr
  \+& \hfill $12.0$
  &&  \hfill $8$
  &&  $0.129913(2)$  \hfill
  &&  \hfill $16$
  &&  $0.129909(2)$    \hfill
  &&\cr
  \+& \hfill $24.0$
  &&  \hfill $8$
  &&  $0.127261(1)$  \hfill
  &&  \hfill $16$
  &&  $0.127258(1)$    \hfill
  &&\cr
\vskip1ex
\thicktablerule
}$$
\endinsert

In table~1 we list the values of $\beta$, the lattice sizes $L/a$ and 
the associated critical hopping parameters $\hopc$
at which the numerical simulations have been performed.
Our final results for the normalization constants are collected in table~2.
For each value of $\beta$ we quote the number obtained 
on the larger lattice with two errors, the first being the 
statistical error, which includes the uncertainty 
in the value of $\hopc$ quoted in table~1.

The second error is an estimate of the systematic effect
which derives from the fact that the chosen lattice sizes 
are not exactly the ones required by the normalization conditions.
The situation at $\beta=6.0$ is exceptional in this respect,
because $L/a=8$ is the correct lattice size 
and the systematic error hence vanishes.
The chosen lattice sizes
$L/a=12$ and $L/a=16$ at $\beta=6.2$ and $\beta=6.4$ 
are rather close to the correct ones. In this case 
the data at $L/a=8$ may be used to estimate the 
change in the normalization constants 
if $L/a$ would be lowered to $11$ and $14$, repectively, which is then 
quoted as the systematic error in table~2.
For $\beta>6.8$ the error is taken to be the difference of 
the normalization constants calculated at $L/a=16$ and $L/a=8$.
Since the effect is of order $(a/L)^2$ 
this procedure appears to be safe and presumably over-estimates
the error.

\vfill\eject

Our numerical results are also shown in fig.~4, where we 
compare them with the one-loop expressions
[\ref{GabrielliEtAl},\ref{GoeckelerEtAl},\ref{StefanNotes}],
$$
  \eqalignno{
  \zv&=1+\zv^{(1)}g_0^2+\rmO(g_0^4),
  \qquad \zv^{(1)}=-0.129430,
  &\enum\cr
  \noalign{\vskip2ex}
  \za&=1+\za^{(1)}g_0^2+\rmO(g_0^4),
  \qquad \za^{(1)}=-0.116458.
  &\enum\cr}
$$
These formulae
describe the data rather well for, say, $g_0^2\leq0.5$, but in 
the range of couplings which is relevant 
for numerical simulations of physically
large lattices this is no longer true.

\topinsert
\newdimen\digitwidth
\setbox0=\hbox{\rm 0}
\digitwidth=\wd0
\catcode`@=\active
\def@{\kern\digitwidth}
\tablecaption{
Results for $\zv$ and $\za$
}
\vskip1ex
$$\vbox{\settabs\+x&xxxxx&%
                  xxxxxxxxx&xxxxxxxxxxxx&%
                  xxxxxxxxx&xxxxxxxxxxxxxx&\cr
\thicktablerule
\vskip1ex
                \+& \hfill     $\beta$ \kern0.3em
                 && \kern2.0em $\zv$   \hfill
                 && \kern2.0em $\za$   \hfill
                 &  \cr
\vskip1.0ex
\thintablerule
\vskip1.5ex
  \+& \hfill $6.0$
  &&  $0.7809(6)$   \hfill
  &&  $0.7906(94)$   \hfill
  &&\cr
  \+& \hfill $6.2$
  &&  $0.7922(4)(9)$   \hfill
  &&  $0.8067(76)(23)$ \hfill
  &&\cr
  \+& \hfill $6.4$
  &&  $0.8032(6)(12)$ \hfill
  &&  $0.8273(78)(10)$   \hfill
  &&\cr
  \+& \hfill $6.8$
  &&  $0.8253(5)(43)$ \hfill
  &&  $0.8549(37)(73)$ \hfill
  &&\cr
  \+& \hfill $7.4$
  &&  $0.8494(3)(34)$  \hfill
  &&  $0.8646(20)(48)$ \hfill
  &&\cr
  \+& \hfill $8.0$
  &&  $0.8667(2)(33)$  \hfill
  &&  $0.8812(19)(17)$ \hfill
  &&\cr
  \+& \hfill $9.6$
  &&  $0.8973(2)(33)$  \hfill
  &&  $0.9078(12)(37)$ \hfill
  &&\cr
  \+& \hfill $12.0$
  &&  $0.9232(2)(26)$  \hfill
  &&  $0.9315(11)(16)$ \hfill
  &&\cr
  \+& \hfill $24.0$
  &&  $0.9656(1)(16)$ \hfill
  &&  $0.9692(4)(14)$ \hfill
  &&\cr
\vskip1ex
\thicktablerule
}$$
\endinsert

The ``mean field improved" perturbation expansion [\ref{Lepenzie}],
$$
  Z=u_0\left\{1+\bigl(Z^{(1)}+\frac{1}{12}\bigr)\gparisi^2+
  \rmO(\gparisi^4)\right\}, 
  \qquad Z=\zv,\za,
  \eqno\enum
$$
comes much closer to the data at low values of $\beta$ (see fig.~4).
The expansion parameter here is
Parisi's boosted bare coupling [\ref{Parisi}],
$$
  \gparisi^2=g_0^2/u_0^4,
  \eqno\enum
$$
with $u_0^4$ being the average plaquette in infinite volume at the
value of $g_0^2$ considered. We mention in passing that 
the data are nearly perfectly matched by the one-loop formulae 
(6.6) and (6.7) if we replace $g_0^2$ by $\gparisi^2$.

\topinsert
\vbox{
\vskip0.0true cm

\centerline{
\epsfxsize=12 true cm
\epsfbox{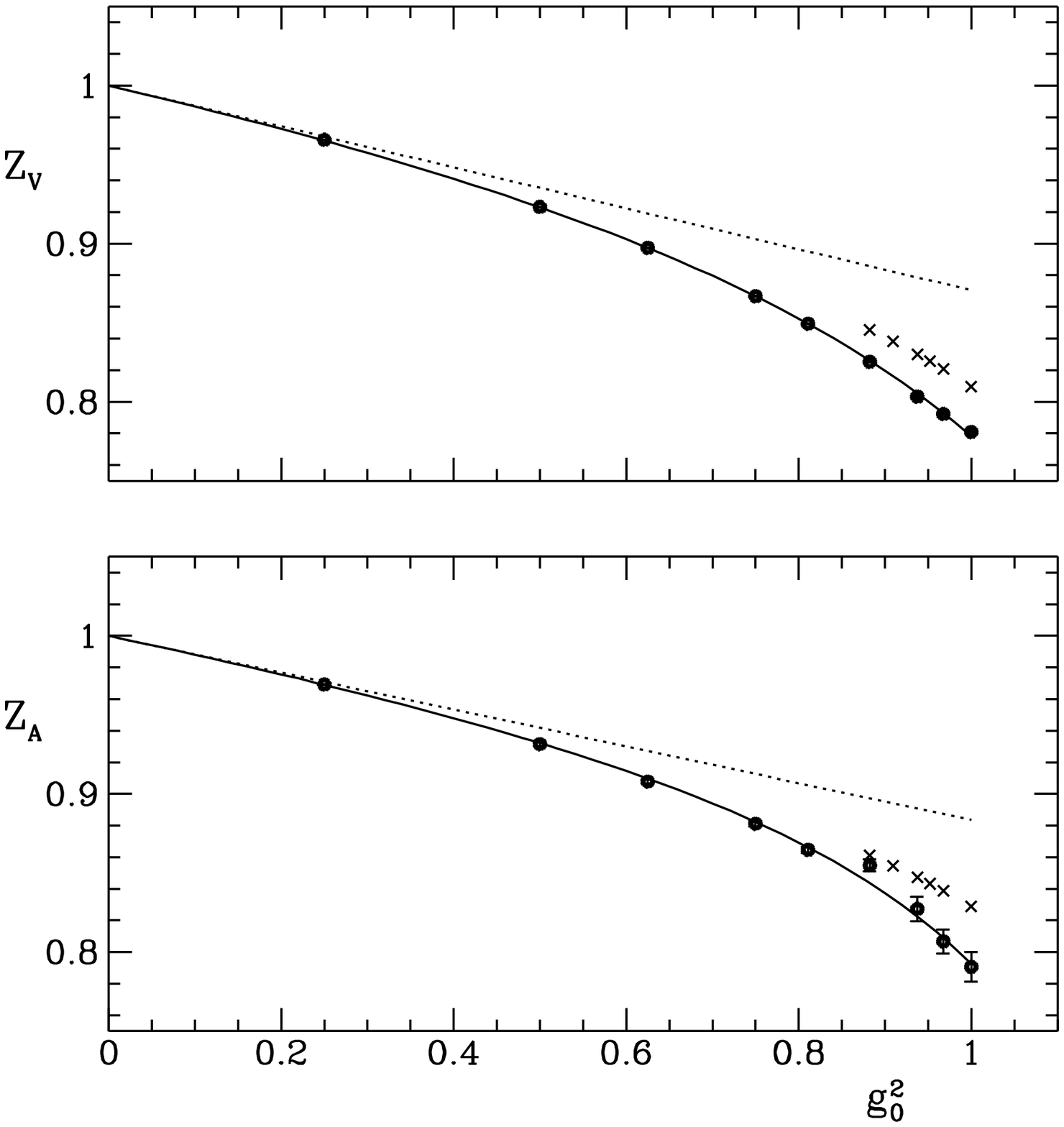}}

\vskip0.0true cm
\figurecaption{Results for $\zv$ and $\za$ from numerical simulations
(filled circles), bare perturbation theory (dotted lines) and ``mean
field improved" perturbation theory (crosses). The solid lines
represent the fits (6.10) and (6.11). For the numerical data only the
statistical errors are displayed.
}
}
\endinsert

In the whole range $0\leq g_0\leq1$
a good representation of the numerical results
is given by the rational expressions
$$
  \eqalignno{
  \zv&={1 - 0.7663 \,g_0^2 + 0.0488 \,g_0^4 \over 1 - 0.6369 \,g_0^2},
  &\enum\cr
  \noalign{\vskip2ex}
  \za&={1 - 0.8496 \,g_0^2 + 0.0610 \,g_0^4 \over 1 - 0.7332 \,g_0^2},
  &\enum\cr}
$$
which coincide with the 
expansions (6.6) and (6.7) to order $g_0^2$. 
The fits reproduce the values of $\zv$ and $\za$ 
quoted in table~2 with a precision better than $0.4\%$ and $0.6\%$,
respectively, an exception being the result for $\za$ at $\beta=6.8$
which deviates by $1.35\%$.
Note, however, that the data points shown in fig.~4 are 
statistically independent and a statistical fluctuation of this
size is hence not an unlikely event.

For future use of our results we suggest to either take 
the numbers quoted in table~2 (where this is possible) or else 
to employ the fit formulae given above, quoting an error of 
$0.5\%$ for $\zv$ and $1.0\%$ for $\za$. 
These error margins should be wide enough to account for 
all the uncertainties in our calculations.

\topinsert
\vbox{
\vskip-2.5true cm

\centerline{
\epsfxsize=11 true cm
\epsfbox{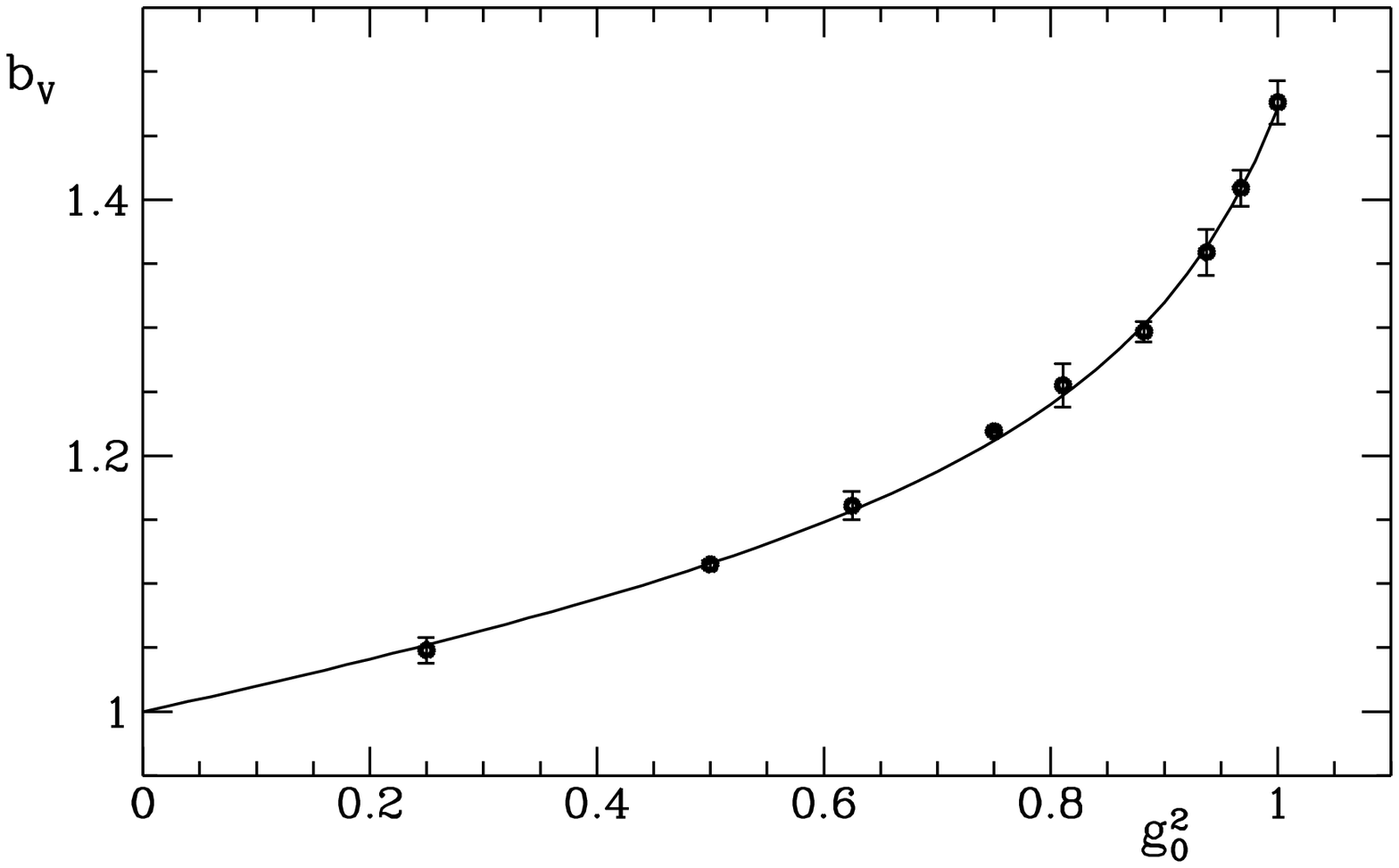}}

\vskip-1.5true cm
\figurecaption{
Numerical results for the coefficient $\bv$.
The solid line represents the fit (6.12).
}
}
\endinsert

\subsection 6.3 Computation of $\bv$

At non-zero quark mass the renormalized improved currents 
involve correction factors of the form $1+\bx a\mq$ that so far are 
only known to lowest order of perturbation theory
[\ref{HeatlieEtAl},\ref{paperII}].
The normalization condition (3.13) for the vector current, 
which is also valid for massive quarks, allows us 
to extract the coefficient $\bv$
by studying the dependence of the ratio $f_1/\fv(x_0)$ on the quark
mass $\mq$. In this subsection we report on our results for $\bv$
obtained on lattices of size $L/a=8$. 
At all couplings $\beta$ considered
we computed the correlation functions at several values of $\hop$
around $\hopc$ and extracted $\bv$ from the linear slope of
$f_1/\fv(T/2)$ as a function of $\hop^{-1}$. 
In the range $0\leq a\mq\leq 0.005$ no 
significant curvature was observed in the data.

The results of our calculations are plotted in fig.~5.
As in the case of the coefficients $\csw$ and
$\ca$, the data can be represented by a global fit
$$
  \bv={{1-0.6518\,g_0^2-0.1226\,g_0^4}\over{1-0.8467\,g_0^2}},
  \qquad 0\leq g_0 \leq1,
  \eqno\enum
$$
which reproduces the tree-level value $\bv=1$ at $g_0=0$.

For studies of pseudoscalar decay constants
it would also be desirable to
know the coefficient $\ba$. The normalization
condition for the axial current derived in subsect.~3.3 is however 
only applicable at zero quark mass.
It is possible to deduce a more general normalization condition
by taking the mass term in the PCAC relation into account,
but the equation that one obtains involves a short distance contribution 
of order $a\mq$ which we would not know how to 
separate from the term proportional to $\ba$.
A more sophisticated approach is hence required to compute $\ba$ 
non-perturbatively.
We should add, however, that a perturbative estimate of 
$\ba$ may be perfectly satisfactory,
if one is interested in situations where $a\mq$ is small
(say less than $0.01$).

\subsection 6.4 Residual cutoff effects in eqs.~(3.13) and (3.18)

Now that the normalization constants 
$\zv$ and $\za$ are known, we can ask how large the 
error term on the right-hand sides
of eqs.~(3.13) and (3.18) is for different choices of the 
kinematical parameters.
In particular,
we can vary the boundary values of the gauge field, 
the angles $\theta_k$ and 
the insertion points $x_0$,~$y_0$, and 
we may also replace one of the 
quarks created and annihilated at the boundaries
of the lattice by a strange quark with non-zero mass
(cf.~subsect.~3.4).
The purpose of such studies is to verify the effectiveness of
O($a$) improvement and also to check that the
particular choices made in subsect.~6.1 do not lead
to uniformly large higher-order cutoff effects in 
other matrix elements of the renormalized currents.

Several tests, covering all variations mentioned
above, have been performed at $\beta=6.4$.
For lattice sizes $L/a\geq8$ we found that 
the cutoff effects would amount to changes in the normalization
constants no larger than the statistical errors quoted in table~2,
thus providing another impressive demonstration of the 
importance and effectiveness of improvement.

\section 7. Concluding remarks

The computation of the current normalization constants $\za$ and $\zv$
complements the non-perturbative determination of the O($a$) counterterms
in the improved action and the improved axial current reported in 
ref.~[\ref{paperIII}]. In particular,
physical matrix elements of the axial current 
in quenched QCD can now be obtained
with O($a$) improvement fully taken into account and
small uncertainty in the normalization factor.
It should again be emphasized, however, that 
an extrapolation to the continuum limit will always be required,
even though we have not observed any
significant residual cutoff effects in the matrix elements considered here.

The methods we have used in this paper carry over literally
to QCD with dynamical quarks.
As a first step one may be interested
in a determination of the normalization constants
using lattices of size $L/a=8$ at all couplings $g_0$.
The results obtained here suggest that the 
associated systematic errors are quite small (at most
$2\%$ in the case of $\za$).
Simulations of larger lattices will however be required
for a reliable estimation of the systematic errors and for
more precise results.

\vskip1ex 
This work is part of the ALPHA collaboration research programme. We
thank DESY for allocating computer time on the APE/Quadrics computers
at DESY-IfH and the staff of the computer centre at Zeuthen for their
support. We are also grateful to Peter Weisz and 
Ulli Wolff for helpful discussions and a critical reading of the paper.
Stefan Sint is partially supported by the U.S.~Department of
Energy (contracts DE-FC05-85ER250000 and DE-FG05-92ER40742). Hartmut
Wittig acknowledges the support of the Particle Physics and Astronomy
Research Council through the award of an Advanced Fellowship.

\appendix A 

We here deduce the Ward identity (3.14) assuming
that the axial current is
conserved (at zero quark mass) and that the operator product
expansion is valid in a weak sense.

The axial current conservation amounts to saying that 
$$
  \langle\partial_{\mu}A^a_{\mu}(x){\cal O}\rangle=0
  \eqno\enum
$$
for any field product $\cal O$ localized in a region not containing $x$.
It follows from this that the integral on the left-hand side of 
eq.~(3.14) is independent of the region $R$ (which must contain 
$y$ and may not have any overlap with the localization 
region of ${\cal O}_{\rm ext}$). 
We may, for example, take $R$ 
to be a ball with small radius $r$ centred at $y$.
For $r\to0$
the integral may then be calculated by inserting the operator
product expansion of $\epsilon^{abc}A^a_{\mu}(x)A^b_{\nu}(y)$. 
Up to logarithmic factors
the contributions 
of the composite fields of dimension $d$ are proportional to $r^{d-3}$.
In particular, fields with dimension $d>3$ make no contribution in the limit
$r\to0$. The only local field with dimension $d\leq3$ 
and the appropriate transformation behaviour under the flavour
and space-time symmetries is the vector current $V^c_{\nu}(y)$.
We thus conclude that eq.~(3.14) must be valid 
up to a proportionality constant $k$.

To prove that $k=1$ we choose $R$ to be the space-time volume between
two equal-time hyper-planes and integrate over the space components
of $y$. For $\nu=0$ the left-hand side of eq. (3.14)
is then equal to some matrix element of 
the commutator of the axial charge with itself, while on the 
other side of the equation one has a matrix element of the isospin
charge between the same states.
With the canonical normalization of the charges the matrix
elements are the same and $k$ must hence be equal to $1$.

\beginbibliography


\bibitem{BochicchioEtAl}
M. Bochicchio et al.,
Nucl. Phys. B262 (1985) 331

\bibitem{MaianiMartinelli}
L. Maiani and G. Martinelli,
Phys. Lett. B178 (1986) 265


\bibitem{MeyerSmith}
B. Meyer and C. Smith, 
Phys. Lett. 123B (1983) 62

\bibitem{MartinelliCheng}
G. Martinelli and Zhang Yi-Cheng,
Phys. Lett. 123B (1983) 433; ibid. 125B (1983) 77

\bibitem{GrootEtAl}
R. Groot, J. Hoek and J. Smit,
Nucl. Phys. B237 (1984) 111


\bibitem{GabrielliEtAl}
E. Gabrielli et al.,
Nucl. Phys. B362 (1991) 475


\bibitem{BorelliEtAl}
A. Borelli, C. Pittori, R. Frezzotti and E. Gabrielli,
Nucl. Phys. B409 (1993) 382


\bibitem{MartinelliEtAlI}
G. Martinelli, S. Petrarca, C. T. Sachrajda and A. Vladikas,
Phys. Lett. B311 (1993) 241, E: B317 (1993) 660

\bibitem{PacielloEtAl}
M. L. Paciello, S. Petrarca, B. Taglienti and A. Vladikas,
Phys. Lett. B341 (1994) 187

\bibitem{HentyEtAl}
D. S. Henty, R. D. Kenway, B. J. Pendleton and J.I. Skullerud,
Phys. Rev. D51 (1995) 5323


\bibitem{MartinelliEtAlII}
G. Martinelli et al.,
Nucl. Phys. (Proc. Suppl.) 42 (1995) 428;
Nucl. Phys. B445 (1995) 81


\bibitem{HeatlieEtAl}
G. Heatlie et al.,
Nucl. Phys. B352 (1991) 266

\bibitem{MartinelliEtAlIII}
G. Martinelli, C. T. Sachrajda and A. Vladikas,
Nucl. Phys. B358 (1991) 212

\bibitem{MartinelliEtAlIV}
G. Martinelli, C. T. Sachrajda, G. Salina and A. Vladikas,
Nucl. Phys. B378 (1992) 591


\bibitem{letter}
K. Jansen et al.,
Phys. Lett. B372 (1996) 275

\bibitem{paperI}
M. L\"uscher, S. Sint, R. Sommer and P. Weisz,
Nucl. Phys. B478 (1996) 365

\bibitem{paperII}
M. L\"uscher and P. Weisz,
Nucl. Phys. B479 (1996) 429

\bibitem{paperIII}
M. L\"uscher, S. Sint, R. Sommer, P. Weisz and U. Wolff,
Non-perturbative O($a$) improvement of lattice QCD,
CERN preprint CERN-TH/96-218 (1996), hep-lat/9609035

\bibitem{BiCGstab}
H. van der Vorst, 
SIAM J. Sc. Comp. 12 (1992) 631

\bibitem{FrommerEtAl}
A. Frommer et al.,
Int. J. Mod. Phys. C5 (1994) 1073

\bibitem{rnull}
R. Sommer,
Nucl. Phys. B411 (1994) 839

\bibitem{HartmutI}
H. Wittig (UKQCD collab.),
Nucl. Phys. B (Proc. Suppl.) 42 (1995) 288;
private notes (1996)


\bibitem{GoeckelerEtAl}
M. G\"ockeler et al.,
Perturbative renormalization of bilinear quark
and gluon operators,
Talk given at the 14th International Symposium on
Lattice Field Theory, St. Louis, 4-8 June 1996,
Humboldt Univ. preprint HUB-EP-96/39, hep-lat/9608033

\bibitem{StefanNotes}
S. Sint, private notes (1996)


\bibitem{Lepenzie}
G. P. Lepage and P. Mackenzie,
Phys. Rev. D48 (1993) 2250

\bibitem{Parisi}
G. Parisi, in: High-Energy Physics --- 1980, XX. Int. Conf.,
Madison (1980), ed. L. Durand and L. G. Pondrom (American Institute
of Physics, New York, 1981)

\endbibliography

\bye